\documentclass[onecolumn,superscriptaddress,showpacs,pra]{revtex4}
\usepackage{epsfig}
\usepackage{epstopdf}
\usepackage{delarray}
\usepackage{amsmath, amssymb}
\usepackage{bm}
\usepackage{graphicx}% Include figure files
\usepackage{placeins}
\usepackage{color}
\begin{document}
\title{Self-Localized Solitons of the Nonlinear Wave Blocking Problem}

\author{Cihan Bay\i nd\i r}
\email{cihanbayindir@gmail.com}
\affiliation{Associate Professor, Engineering Faculty, \.{I}stanbul Technical University, 34469 Maslak, \.{I}stanbul, Turkey. \\
						 Adjunct Professor, Engineering Faculty, Bo\u{g}azi\c{c}i University, 34342 Bebek, \.{I}stanbul, Turkey. \\
						 International Collaboration Board Member, CERN, CH-1211 Geneva 23, Switzerland.}

%\date{\today}
\begin{abstract}
In this paper, we propose a numerical framework to study the shapes, dynamics and the stabilities of the self-localized solutions of the nonlinear wave blocking problem. With this motivation, we use the nonlinear Schr\"{o}dinger equation (NLSE) derived by Smith as a model for the nonlinear wave blocking. We propose a spectral renormalization method (SRM) to find the self-localized solitons of this model. We show that for constant, linearly varying or sinusoidal current gradient, i.e. $dU/dx$, the self-localized solitons of the Smith's NLSE do exist. Additionally, we propose a spectral scheme with $4^{th}$ order Runge-Kutta time integrator to study the temporal dynamics and stabilities of such solitons. We observe that self-localized solitons are stable for the cases of constant or linearly varying current gradient  however, they are unstable for sinusoidal current gradient, at least for the selected parameters. We comment on our findings and discuss the importance and the applicability of the proposed approach.

%\pacs{03.65.−w, 05.45.-a, 03.75.−b}
\end{abstract}
\maketitle

%%%%%%%%%%%%%%%%%%%%%%%%%%%%%%% main %%%%%%%%%%%%%%%%%%%%%%%%%%%%%
\section{Introduction}
The phenomenon of wave blocking can be described as the stopping of the propagating waves by opposing ocean currents and circulations. The celerity of waves propagating into the opposing oceanic currents and circulations reduces and if opposing current is strong enough, then they may be blocked, that is the group velocity of the wavefield becomes zero. This phenomena is observed in the nature. One of the most common places of their observation is tidal inlets where tidal currents are strong enough to block the propagating waves. Due to the sharp increase in wave steepness during wave blocking processes, the wave environment tends to have higher waves which eventually may become rogue and they may break. Therefore, wave blocking may result in catastrophic navigational hazard in the marine environment. The boats have been known to capsize during crossing inlets, where wave blocking phenomena are observed. Additionally, due to blocking of the waves by Agulhas current, the South Africa off coast is known as one of the most dangerous seas of the earth \cite{Kharif}.

With these motivations, scientists have spent considerable effort to study the wave blocking problem using alternative approaches. Spectral modeling of current induced wave blocking is given in \cite{Ris}. Reflection of short gravity waves on a non-uniform current is studied in \cite{Shyu3}. The blockage of gravity and capillary waves by longer waves and currents are analyzed in \cite{Shyu1}. Reflection of oblique waves by currents are analyzed analytically and numerically in \cite{Shyu2}. Using the wave focusing and hydraulic jump formation approach discussed in \cite{Chin1, Chin2, Chin3}, the kinematic barrier for gravity waves on variable current is discussed in \cite{Chin4}. Monochromatic and random wave breaking at blocking points is studied in \cite{Chawla2}. A model for blocking of periodic waves is proposed in \cite{Suastika1, Suastika2}. Additionally, some of the experimental studies on wave blocking can be seen in \cite{Lai, Pokazayev, Chawla1}. While majority of these studies analyzes wave blocking phenomena within the frame of linear theory, some studies utilizing the nonlinear theory also exist \cite{Smith1976, Peregrine, Chawla3, BayISOPE}.

In this paper, we follow a different methodology for analyzing the blocked solitons of the nonlinear wave blocking problem. We use an extended version of the nonlinear Schr\"{o}dinger equation (NLSE), first derived in \cite{Smith1976} as the nonlinear wave-current interaction model. This equation can be named as Smith's NLSE. We develop a spectral renormalization method (SRM) for the numerical solutions of the Smith's NLSE which can be implemented for arbitrary current gradient forms, i.e. $dU/dx$. This SRM can be used to find the self-localized solitons starting from an arbitrary wave profile. We show that self-localized solutions of the nonlinear wave blocking problem can be found using the approach proposed in this paper. Additionally, we perform the stability analysis for the solitons blocked by oceanic current. The stability analysis includes an assessment of the Vakhitov-Kolokolov slope condition and the time stepping of the solitons found by SRM. We implement a Fourier spectral scheme with $4^{th}$ order Runge-Kutta time integrator to analyze the temporal dynamics and stabilities of the self-localized solitons, for the time stepping part of the stability problem. We show that self-localized solitons are stable for the cases of constant or linearly varying current gradient, at least for selected parameters.  However, our results also indicate that they are unstable for sinusoidal current gradient, $dU/dx$, at least for the parameters considered. We discuss the uses, applicability and limitations of the proposed approach and comment on our findings.

\section{\label{sec:level2}Methodology}

\subsection{\label{sec:level1} Review of the Nonlinear Schrödinger Equation for Wave Blocking}
Various models have been proposed to study the wave blocking phenomena. In majority of these studies have blocking is studied within the frame of the linear theory. However, there are some studies which utilizes nonlinear theories with this aim \cite{Smith1976, Peregrine, Chawla3, BayISOPE}. Smith \cite{Smith1976} derived the formula
\begin{equation}
i\psi_t - \frac{\omega}{8k^2} \psi_{xx} - \frac{\omega k^2}{2} \left|\psi \right|^2 \psi + k\left|\frac{dU}{dx} \right|x\psi =0
\label{eq01}
\end{equation}
to study the nonlinear wave field under the effect of wave blocking. This formula is valid in the vicinity of blocking point due to Taylor series expansion used in its derivation \cite{Kharif, Smith1976}. In here, $x, t$ denote the spatial and temporal variables, $i$ denotes the imaginary unity, $\psi$ is complex amplitude which represents the envelope of the wavefield, $k$ is the wavenumber which is selected as $k=0.05 rad/m$ throughout this study, $\omega$ is the wave frequency, $U$ is the current speed and $dU/dx$ is the current gradient.  The last term in Eq.~(\ref{eq01}) accounts the Doppler shift of the wave frequency by current. If one neglects the nonlinearity, then Eq.~(\ref{eq01}) reduces to the Airy equation studied in \cite{Suastika1, Smith1976}. Setting the group velocity to zero, that is
\begin{equation}
C_g=\frac{d\omega}{dk}= \frac{1}{2}\sqrt{\frac{g}{k}} + U(x_o)=0
\label{eq02}
\end{equation}
the blocking current speed and blocking point can be found. In here, $C_g$ is the group velocity,  $x_o$ is the blocking and $g$ is the gravitational acceleration. The angular frequency, $\omega$, can be computed by
\begin{equation}
\omega(k)=\pm \sqrt{gk} + k U(x_o)
\label{eq03}
\end{equation}
Throughout this study the negative root is used. It is useful to note that, when the $\left|dU/dx \right|$ term is a constant, then a simple transformation in the form   
\begin{equation}
\psi(x,t)= \widetilde{\psi} \left (x+\frac{t^2}{2}\left|\frac{dU}{dx} \right|, t \right) \exp{\left( -i \left[xt \left|\frac{dU}{dx} \right|+\frac{t^3}{6} \left|\frac{dU}{dx} \right|^2 \right] \right) }
\label{eq04}
\end{equation}
transforms the Smith's NLSE to the standard cubic NLSE \cite{ChenLiu}, given as
\begin{equation}
i\widetilde{\psi}_t + \frac{1}{2} \widetilde{\psi}_{xx} +  \left|\widetilde{\psi} \right|^2 \widetilde{\psi} =0
\label{eq05}
\end{equation}
Therefore, it becomes obvious that stable one, two \cite{Kharif}, N-solitons \cite{BayRINP}, rogue waves \cite{Kharif, BayPLA, BayPRE1, BayPRE2}, and other type of waves can be found for the nonlinear wave blocking problem if the term $\left|dU/dx \right|$ is a constant. The solitons of this model move towards the blocking point, get blocked by the current at blocking point and then they are reflected  \cite{Kharif}. Throughout this process soliton amplitude remains constant due to the balance between defocusing (dispersive focusing) and amplification (attenuation) by the non-uniform current \cite{Kharif}. Our paper aims to overcome the limitations on the form of the current gradient. In the case of variable current gradients, i.e. $dU/dx$ term, the self-localized solutions of nonlinear wave blocking problem can be found and analyzed using the SRM discussed in the next section.

\subsection{Spectral Renormalization Method for the Smith's Nonlinear Schrödinger Equation} 

There are few different computational and analytical techniques which can be used to find the self-localized solutions of various nonlinear equations \cite{Ablowitz}. Shooting, self-consistency and relaxation techniques are some of them \cite{Ablowitz, Bay_CSRM}. Petviashvili's method (PM) is another technique commonly used for the same purpose. In PM, as in the case of general Fourier spectral schemes \cite{Canuto, trefethen,Baysci}, the model nonlinear equation is transformed into wavenumber space. Then, depending on the degree of nonlinearity a convergence factor is determined  \cite{Ablowitz, Petviashvili}. PM was first applied to the Kadomtsev-Petviashvili equation \cite{Petviashvili}, then it has been generalized to many different equations to model many different phenomena including but are not limited to lattice vortices, gray and dark solitons \cite{BayRINP, Ablowitz, Bay_CSRM}. Although PM performs well for nonlinearities with fixed homogeneity, there are many different systems with different types of nonlinearity thus an extension of PM which is known as spectral renormalization method (SRM) is proposed in \cite{Ablowitz, Fibich}. In SRM, the nonlinear governing equation is transformed into wavenumber space by means of Fourier transform. This transformed equation is coupled with a nonlinear integral equation, which is basically an energy conservation equation for the iterations in the wavenumber space \cite{Ablowitz}. This coupling ensures that the initial conditions converge to the self-localized solutions of the model equation under investigation \cite{Ablowitz}. SRM is an efficient, easy to implement technique. Thus, it is applied to many different nonlinear models with different nonlinearity types \cite{BayRINP, Ablowitz}. Although SRM is quite well known by the researchers in the fields of fiber optics, it is not very well known in the ocean modelling community. We propose a SRM to the nonlinear wave blocking problem in this paper, which also aims to do this introduction. 

Smith's NLSE given by Eq.~(\ref{eq01}) is
\begin{equation}
i\psi_t - \frac{\omega}{8k^2} \psi_{xx} - \frac{\omega k^2}{2} \left|\psi \right|^2 \psi + k\left|\frac{dU}{dx} \right|x\psi =0
\label{eq06}
\end{equation}
which can be rewritten as
\begin{equation}
i\psi_t - \frac{\omega}{8k^2} \psi_{xx} +V(x)\psi+ N(\left| \psi \right|^2) \psi =0
\label{eq07}
\end{equation}
where $N(\left| \psi \right|^2)=- \frac{\omega k^2}{2} \left|\psi \right|^2$ and $V(x)=k\left|\frac{dU}{dx} \right|x$ is the potential term. Pluging in the ansatz, $\psi(x,t)=\eta(x,\mu) \textnormal{exp}(i\mu t)$, where $\mu$ represents the soliton eigenvalue, the Smith's NLSE becomes
\begin{equation}
-\mu \eta - \frac{\omega}{8k^2}  \eta_{xx} +V(x)\eta+ N(\left| \eta \right|^2) \eta =0
\label{eq09}
\end{equation}
The 1D Fourier transforming of $\eta$ can be calculated as
\begin{equation}
\widehat{\eta} (m)=F[\eta(x)] = \int_{-\infty}^{+\infty} \eta(x) \exp[i(mx)]dx
\label{eq10}
\end{equation}
where $m$ is the Fourier transform parameter. By taking the Fourier transform of the Eq.~(\ref{eq09}), one can obtain the iteration scheme. However it is known that this scheme exhibits singularities \cite{Ablowitz}. In order to avoid singularity of the scheme, we add and substract a $p \eta$ term from the 1D Fourier transform of Eq.~(\ref{eq09}) \cite{Ablowitz}. Then the iteration scheme becomes
\begin{equation}
\widehat{\eta} (m)=\frac{(p+| \mu|)\widehat{\eta}}{p- \frac{\omega}{8k^2} \left| m \right|^2} +\frac{F[V \eta]} {p- \frac{\omega}{8k^2} \left| m \right|^2}+\frac{F \left[ N(\left| \eta \right|^2) \eta \right]}{p- \frac{\omega}{8k^2} \left| m \right|^2}
\label{eq11}
\end{equation}
In here, since $\frac{\omega}{8k^2}<0$ for the selected values of $k$ and $\omega$, a constant $p=-10$ satisfying $p<0$ is used to avoid singularity. It is known that these iterations may not converge, they either may grow unboundedly or they may tend to zero as \cite{Ablowitz}. One can overcome this problem by introducing a new variable, $\eta(x)=\alpha \xi(x)$. Thus, the 1D Fourier transform of this new variable becomes $\widehat{\eta}(m)=\alpha \widehat{\xi}(m)$. Using these substitutions, Eq.~(\ref{eq11}) becomes
\begin{equation}
\widehat{\xi}_{j+1} (m) =\frac{(p+|\mu|)}{p-\frac{\omega}{8k^2} \left| m \right|^2}\widehat{\xi_j}+\frac{F[V \xi_j]}{p -\frac{\omega}{8k^2}\left| m \right|^2}
+\frac{F\left[  - \frac{\omega k^2}{2} \left| \alpha_j \right|^2 \left|\xi_j  \right|^2 \xi_j \right]}{p-\frac{\omega}{8k^2}\left| m \right|^2} =R_{\alpha_j}[\widehat{\xi}_j (m)]
\label{eq12}
\end{equation}
which becomes the iteration scheme. It is possible to attain the algebraic condition for SRM by multiplying both sides of Eq.~(\ref{eq12}) with  $\widehat{\xi}^*(m)$ and integrating to calculate the energy. This integration gives the normalization constraint as
\begin{equation}
\int_{-\infty}^{+\infty} \left|\widehat{\xi} (m)\right|^2 dm= \int_{-\infty}^{+\infty} \widehat{\xi}^* (m) R_{\alpha}[\widehat{\xi} (m)]dm  
\label{eq13}
\end{equation}
At this point it is possible to summarize the SRM; an initial condition given in the form of a single or multi-Gaussians, converges to self-localized states of the model equation when Eq.~(\ref{eq12}) and Eq.~(\ref{eq13}) are iteratively applied. Once the parameter ${\alpha}$ converges to a specified upper error bound, the iterations can be ceased. A more comprehensive discussion of SRM can be seen in \cite{Ablowitz}.

The self-localized solutions can be found for the Smith's NLSE using the SRM summarized above. However, it is also very important to discuss the stability characteristics of those solutions. There are two conditions for the soliton stability. The first condition is the Vakhitov and Kolokolov slope condition. This necessary condition states that, $dP/d\mu <0$ should be satisfied. In here $P=\int \left| \psi  \right|^2 dx$ is the soliton power and $\mu$ is the soliton eigenvalue \cite{VakhitovStability, WeinsteinStability, SivanStability}. The second condition for the soliton stability is the spectral condition, which is sometimes studied by analyzing the eigenvalues of the operator of the governing equation or more commonly by a numerical approach \cite{WeinsteinStability, SivanStability}. In the coming sections we present the calculations for the first stability condition. For the second stability condition, we implement a Fourier spectral scheme with $4^{th}$ order Runge-Kutta time integrator for time stepping the Smith's NLSE. We summarize this numerical scheme below.

One can rewrite the Smith's NLSE given by Eq.(\ref{eq01}) as
\begin{equation}
\psi_t =-i \left( \frac{\omega}{8k^2} \psi_{xx} + \frac{\omega k^2}{2} \left|\psi \right|^2 \psi - k\left|\frac{dU}{dx} \right|x\psi \right) =h(\psi, t,x)
\label{eq014}
\end{equation}
where the RHS is denoted by a new function, $h(\psi, t,x)$. The second order derivative in this formula can be calculated using the Fourier series at each time step as
\begin{equation}
\psi_{xx} = F^{-1} \left[ -m^2 F[\psi] \right]
\label{eq015}
\end{equation}
where $F$ and $F^{-1}$ shows the Fourier and the inverse Fourier transforms, respectively. In here, $m$ is the wavenumber vector. This wavenumber vector has exact $N$ multiples of the fundamental wavenumber, $m_o=2 \pi/L$. The domain length is selected to be $L=100$ and the number of spectral components is selected to be $N=1024$ for the efficient computations of FFTs. The nonlinear products are calculated by simple multiplication in the spatial domain. The four slopes of the $4^{th}$ order Runge-Kutta scheme can be calculated as
\begin{equation}
\begin{split}
& s_1=h(\psi^n, t^n, x) \\
& s_2=h(\psi^n+0.5 s_1dt, t^n+0.5dt, x) \\
& s_3=h(\psi^n+0.5 s_2dt, t^n+0.5dt, x) \\
& s_4=h(\psi^n+s_1dt, t^n+dt, x) \\
\end{split}
\label{eq016}
\end{equation}
where $dt$ is the time step which is selected as $dt=5 \times 10^{-4}$ throughout in this study. Then the values of the complex wavefunction and time at the next time steps can be found using the expressions
\begin{equation}
\begin{split}
& \psi^{n+1}=\psi^{n}+(s_1+2s_2+2s_3+s_4)/6 \\
& t^{n+1}=t^n+dt\\
\end{split}
\label{eq017}
\end{equation}
iteratively, where n is the iteration count. The initial conditions of this scheme is the self-localized solitons found by SRM. By keeping the soliton power information as a function of time, it is possible to study the stability characteristics of the self-localized solitons.

\section{\label{sec:level3}Results and Discussion}

\subsection{Self-Localized Solitons For Constant Current Velocity Gradient, dU/dx=Constant Case.}

\begin{figure}[htb!]
\begin{center}
   \includegraphics[width=3.4in]{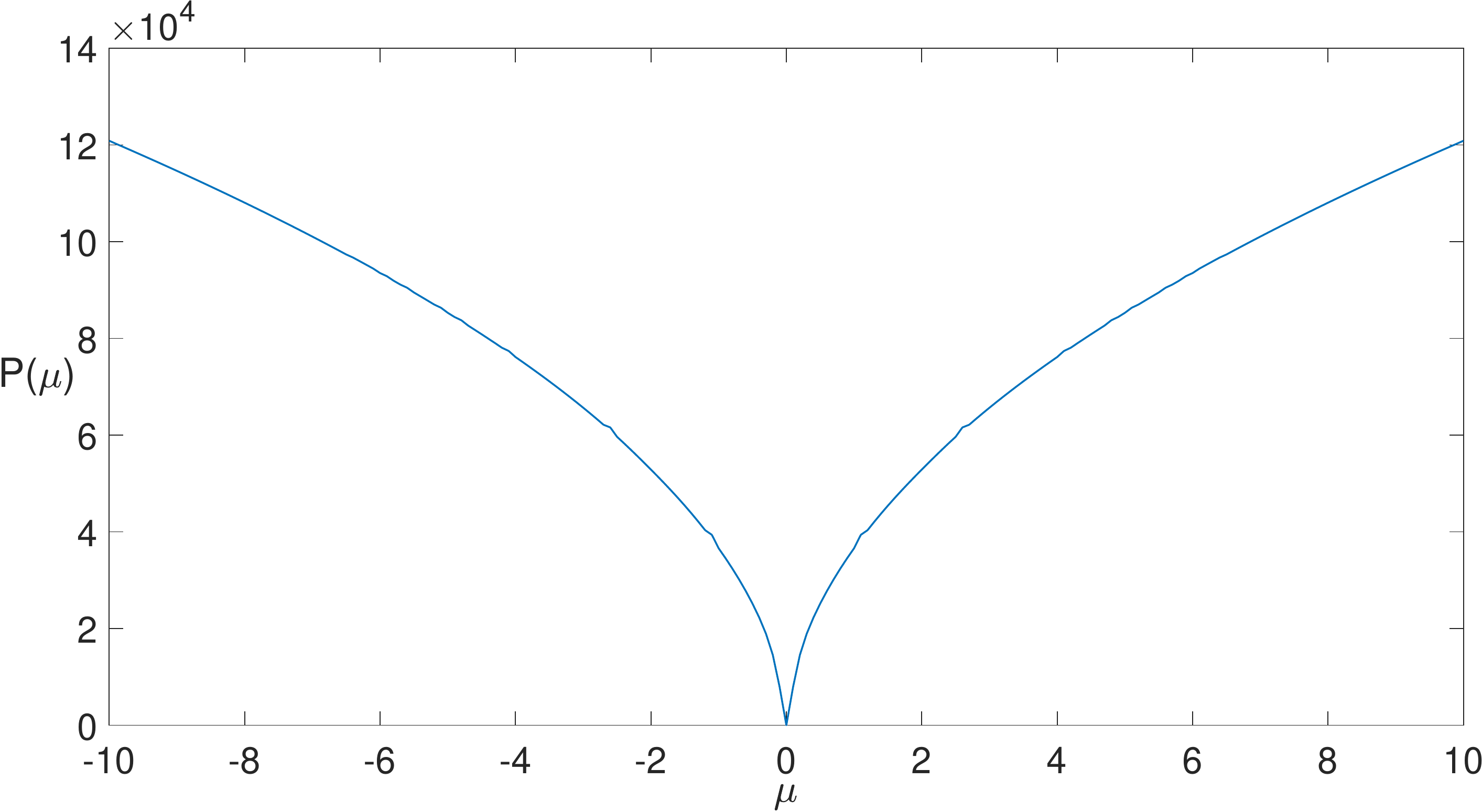}
  \end{center}
\caption{\small Self-localized soliton power as a function of soliton eigenvalue $\mu$ for $dU/dx=0.1 s^{-1}$.}
  \label{fig1}
\end{figure}

In this section, we consider the case of constant current velocity gradient and select $dU/dx=0.1 s^{-1}$ in our simulations. The initial condition for this case is selected to be a Gaussian in the form of $\exp{(-(x-x_0)^2)}$, where $x_0$ is taken as $0$. We define the convergence as the normalized change of $\alpha$ to be less than $10^{-7}$ in our iterations. In Fig.~\ref{fig1}, we plot the self-localized single soliton power as a function of soliton eigenvalue, which is obtained by using the SRM summarized above for different values of soliton eigenvalue, $\mu$.

\begin{figure}[htb!]
\begin{center}
   \includegraphics[width=3.4in]{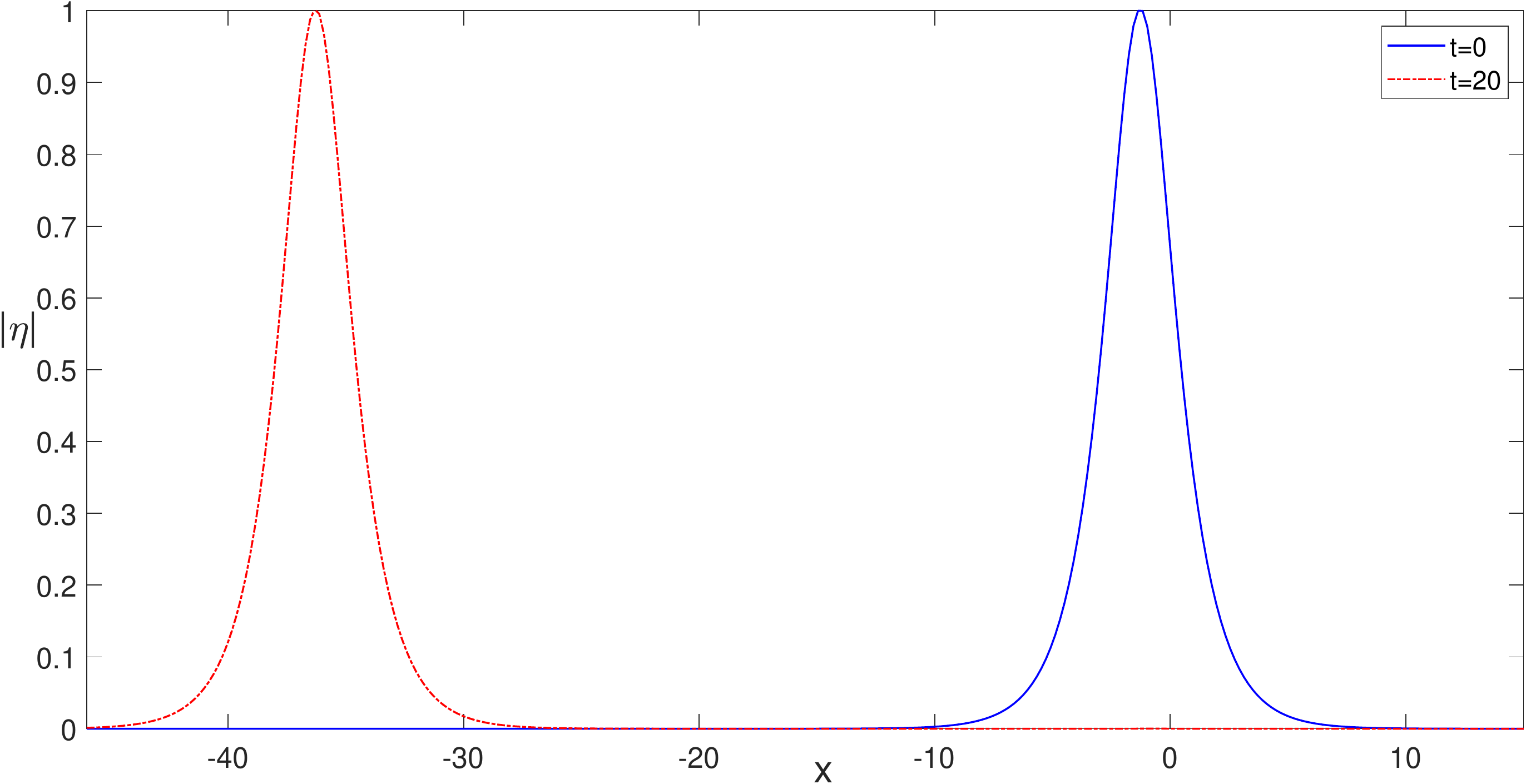}
  \end{center}
\caption{\small Self-localized soliton at times $t=0s$ and $t=20s$ for $dU/dx=0.1 s^{-1}$.}
  \label{fig2}
\end{figure}

Checking Fig.~\ref{fig1}, one can realize that necessary Vakhitov-Kolokolov slope condition for slope stability, i.e. $dP/d\mu <0$ is satisfied by negative values of $\mu$, thus a value of $\mu=-10$ is used. The initial Gaussian profile converges to the exact single sech type self-localized soliton given in \cite{Kharif} within few iteration steps of SRM. This self-localized soliton is depicted in Fig.~\ref{fig2} at two different times, i.e. $t=0s$ and $t=20s$.

\begin{figure}[htb!]
\begin{center}
   \includegraphics[width=3.4in]{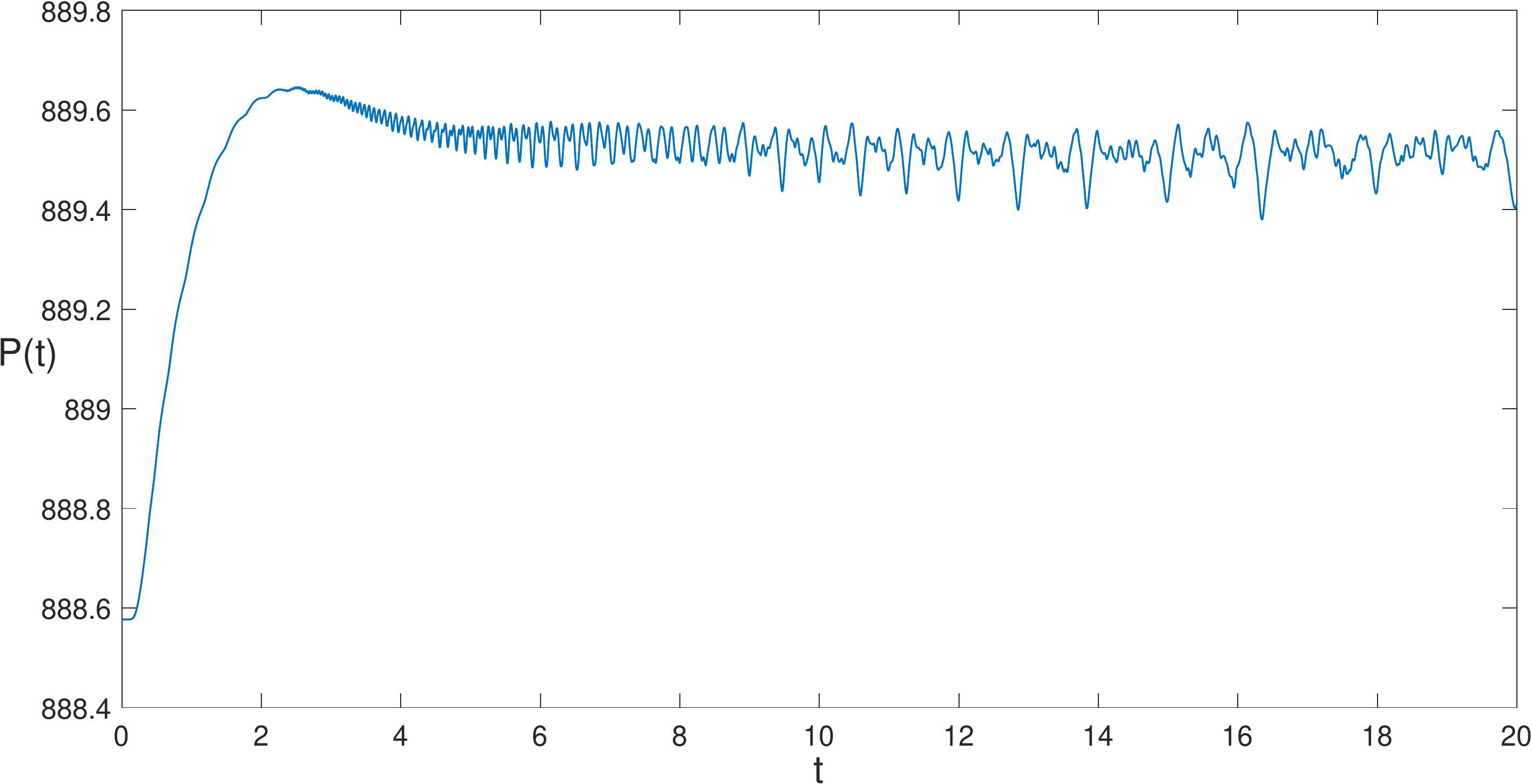}
  \end{center}
\caption{\small Self-localized soliton power as a function of time for $dU/dx=0.1 s^{-1}$.}
  \label{fig3}
\end{figure}

Checking Fig.~\ref{fig2}, it is possible to observe that the self-localized soliton is reflected by the current as expected and discussed in \cite{Kharif}. Checking Fig.~\ref{fig2} and Fig.~\ref{fig3}, one can argue that the soliton shape and amplitude is preserved during temporal evolution, and soliton power remains bounded in time. Thus one can argue that the soliton under constant current velocity gradient of $dU/dx=0.1 s^{-1}$ is stable, at least for the time range considered. 

Additionally, we observe that when the simulations are initiated with two or more Gaussians (N-Gaussians), it is possible to find the two self-localized solitons or N-self localized solitons, in general. However, the convergence criteria should be relaxed for finding those solutions. A convergence criteria of the normalized change of $\alpha$ to be less than $10^{-5}$ can be used to find two self-localized solitons and to be less than $10^{-2}$ can be used to find three soliton solutions. We additionally observe that, during time stepping, the soliton power remains bounded for two soliton case, thus it can be argued that they are stable. For the sake of brevity of the presentation, we turn our attention to other current velocity gradient cases. 

\subsection{Self-Localized Solitons For Linearly Varying Current Velocity Gradient, dU/dx=Linearly Varying Case.}

Next, we turn our attention to the case of linearly varying current velocity gradient and select $dU/dx=0.005x \ s^{-1}$ in the simulations performed in this part. The initial condition for this case is again a Gaussian in the form of $\exp{(-(x-x_0)^2)}$, where $x_0$ is as before. Similarly, the convergence factor is defined as the normalized change of $\alpha$ to be less than $10^{-7}$. The soliton power as a function of soliton eigenvalue for this case closely follows the behavior depicted in Fig.~\ref{fig1}, thus for the sake of brevity is not repeated here. As before, for the soliton eigenvalue a value of $\mu=-10$ is used since the Vakhitov-Kolokolov stability condition is satisfied for this value.

\begin{figure}[htb!]
\begin{center}
   \includegraphics[width=3.4in]{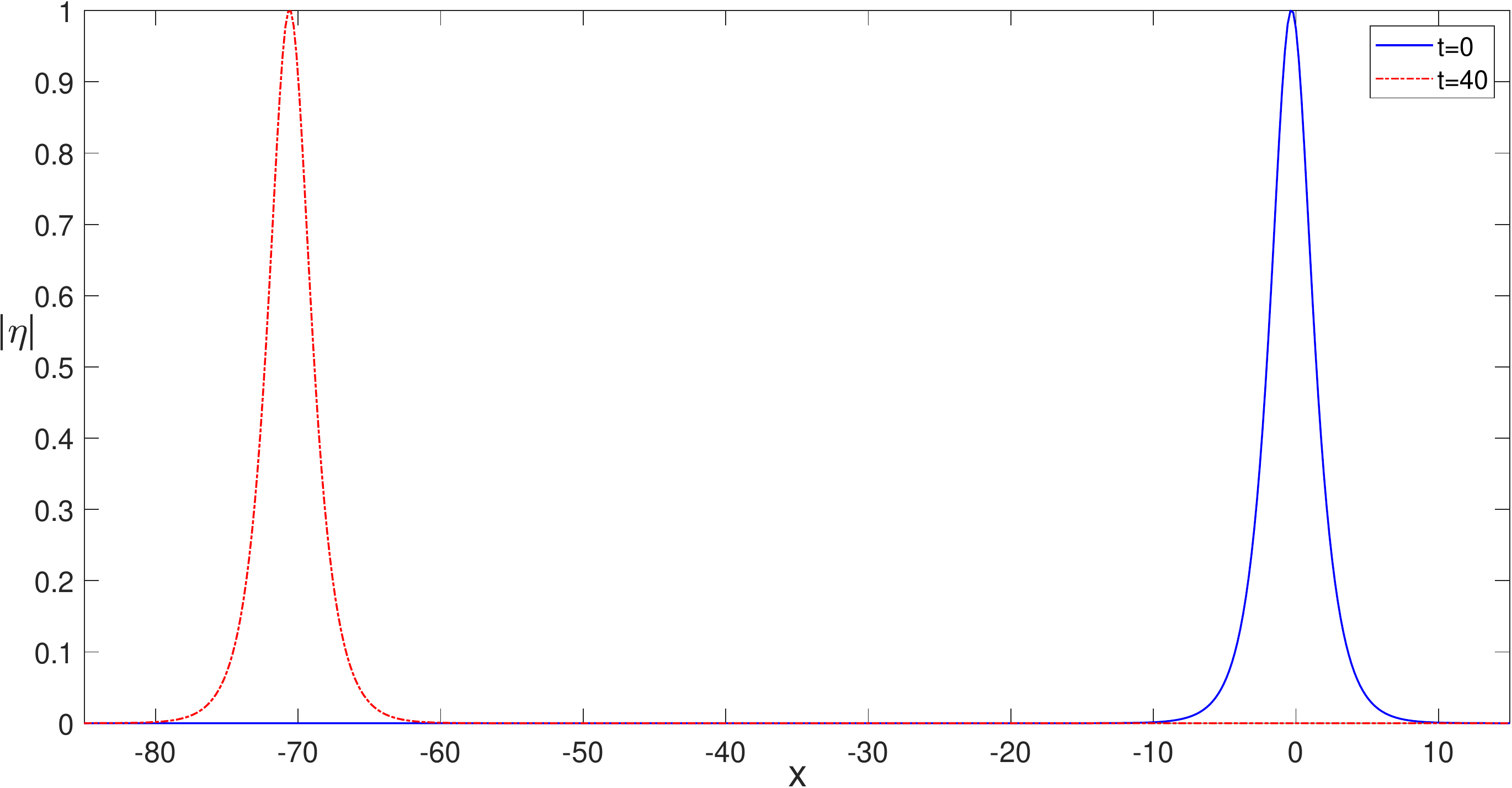}
  \end{center}
\caption{\small Self-localized soliton at times $t=0s$ and $t=40s$ for $dU/dx=0.005x \ s^{-1}$.}
  \label{fig4}
\end{figure}

\begin{figure}[htb!]
\begin{center}
   \includegraphics[width=3.4in]{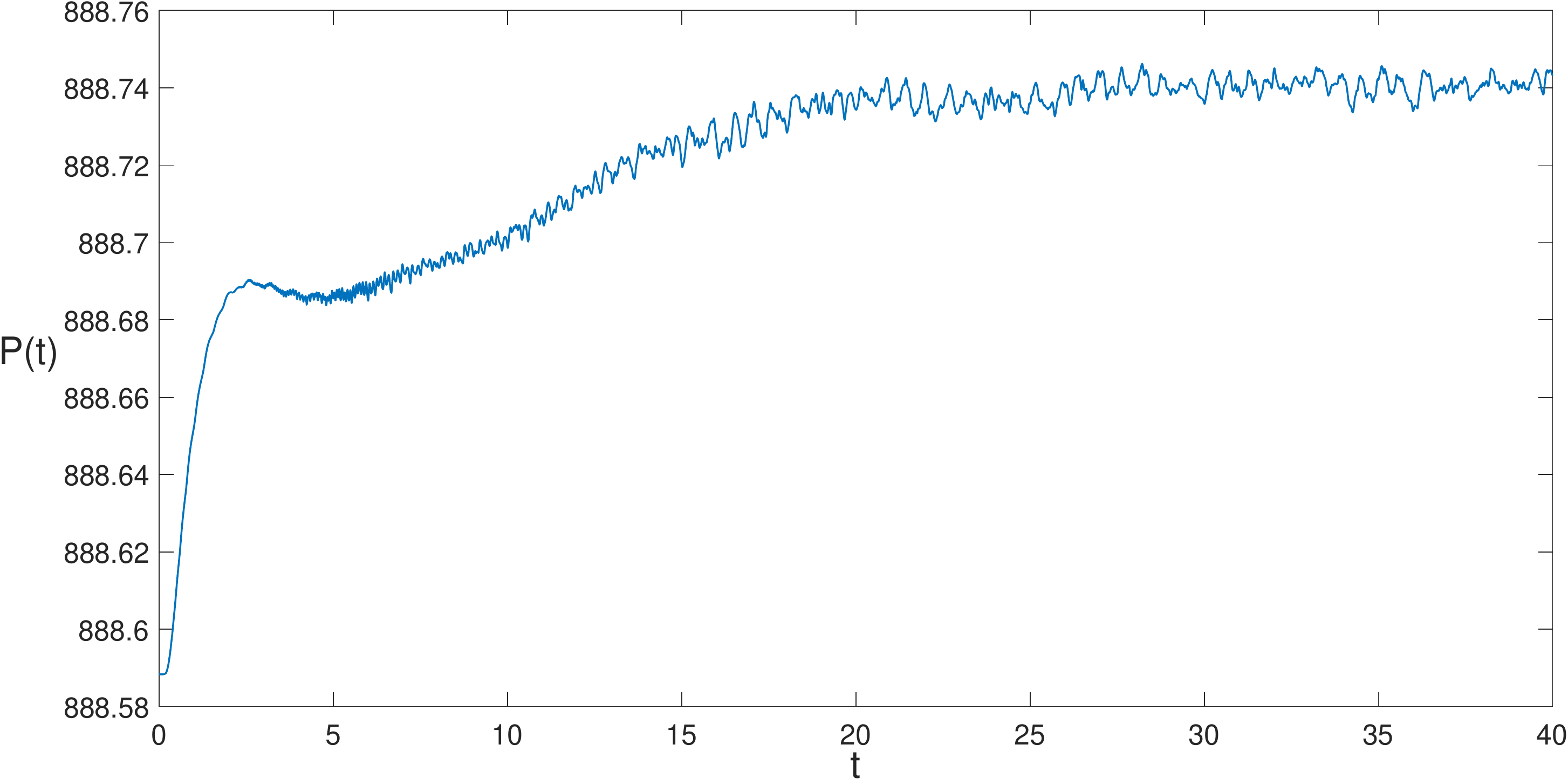}
  \end{center}
\caption{\small Self-localized soliton power as a function of time for $dU/dx=0.005x \ s^{-1}$.}
  \label{fig5}
\end{figure}

The SRM for this case converges to the soliton depicted in Fig.~\ref{fig4} at two different times of $t=0s$ and $t=40s$. Checking Fig.~\ref{fig4}, one can also see that soliton shape is preserved during time stepping while to soliton is reflected by the current leftwards. As depicted in Fig.~\ref{fig5}, it is possible to argue that soliton power is stabilized after a longer initiation time and then remains bounded. We observe a similar tendency for the two soliton found, when two Gaussians are used as initial conditions in the SRM.

\subsection{Self-Localized Solitons For Sinusoidally Varying Current Velocity Gradient, dU/dx=Sinusoidally Varying Case.}
\begin{figure}[htb!]
\begin{center}
   \includegraphics[width=3.4in]{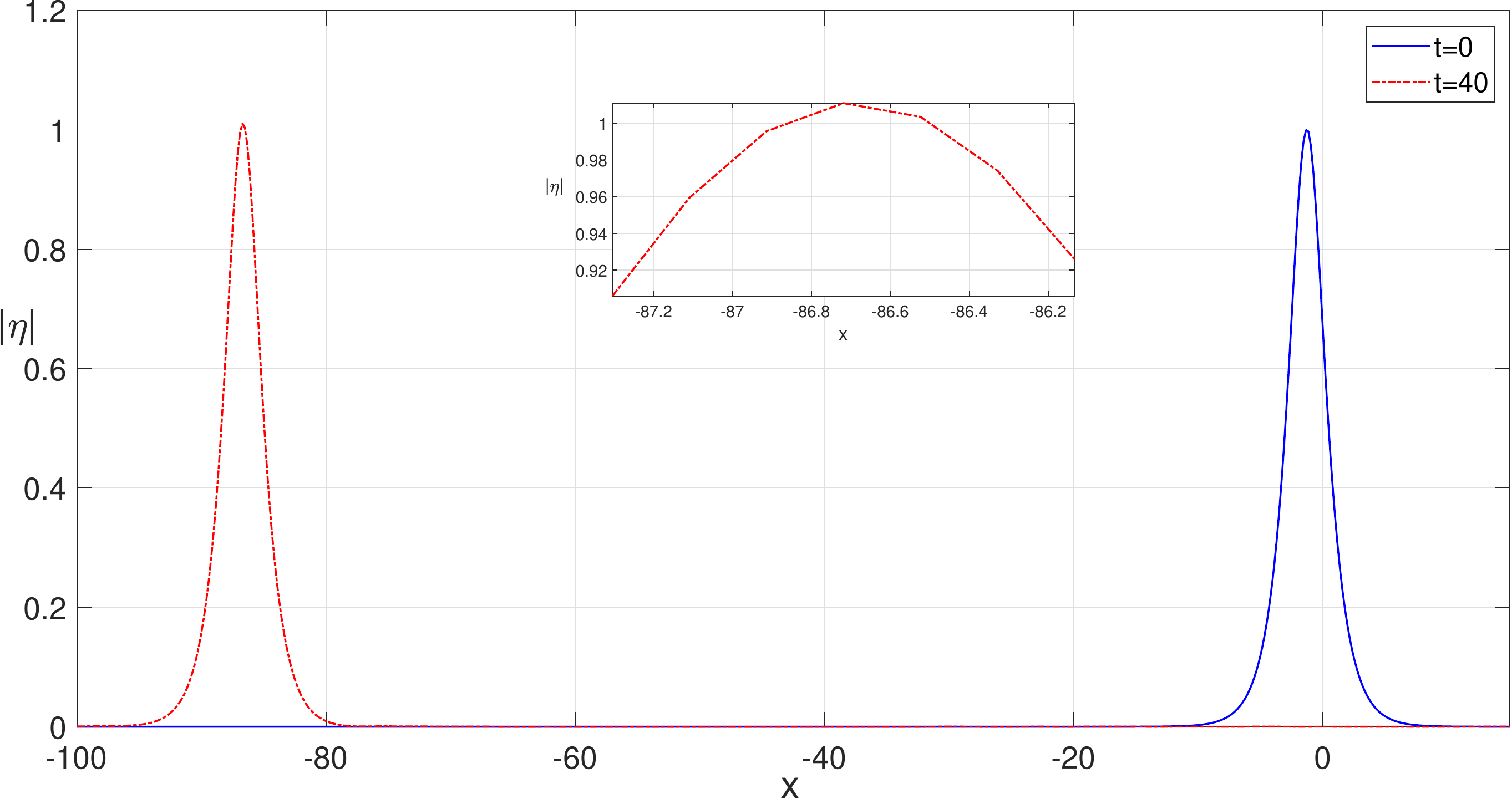}
  \end{center}
\caption{\small Self-localized soliton at times $t=0s$ and $t=40s$ for $dU/dx=0.1sin(x) \ s^{-1}$.}
  \label{fig6}
\end{figure}

Lastly, we analyze the case of sinusoidally varying current velocity gradient and select $dU/dx=0.1sin(x) \ s^{-1}$ in the simulations performed in this part of the paper. As before, when the same parameters are used in SRM iterations, an initial condition in the form of a Gaussian converges to the soliton depicted in Fig.~\ref{fig6} at two different times of of $t=0s$ and $t=40s$.

\begin{figure}[htb!]
\begin{center}
   \includegraphics[width=3.4in]{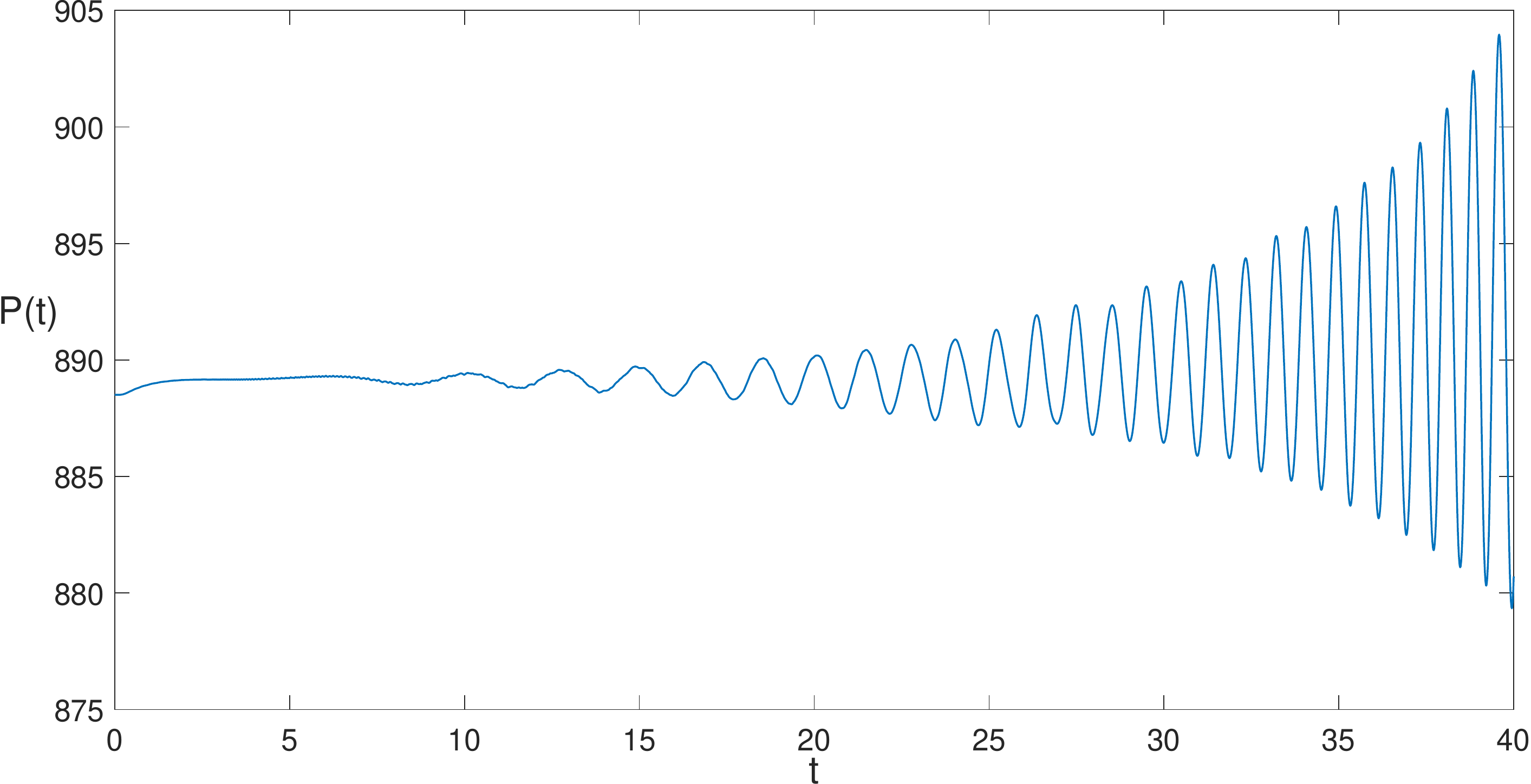}
  \end{center}
\caption{\small Self-localized soliton power as a function of time for $dU/dx=0.1sin(x) \ s^{-1}$.}
  \label{fig7}
\end{figure}

A careful analysis of Fig.~\ref{fig6} and the time series of soliton power given in Fig.~\ref{fig7} indicate that the self-localized soliton is unstable for the case of sinusoidally varying current velocity gradient, at least for the parameter values considered. We observe a similar tendency for self-localized two soliton blocked by sinusoidally varying ocean current gradient, which eventually turns out to be unstable.

It is important to emphasize then starting from various forms of initial conditions, the SRM converges to self-localized soliton in the form of single or multiple sech functions. Thus, arbitrary wave profiles can be used as initial conditions in the SRM to discuss their stability characteristics and possible evolutions to self-localized profiles. In the literature, there are many extended versions of the NLSE. For example, Smith's NLSE does not take the self-steepening, wave skewness, Raman scattering or quintic nonlinearity effects into account. However, the procedure given in our paper can be easily generalized to analyze the blocking, steepening and breaking properties of solitons under such effects. However, blocking properties of different wave types such as the rogue waves can not be analyzed using the presented form of the SRM, since it does not converge to rational solutions of the model investigated but it  converges to the self-localized solitons.

\section{\label{sec:level1}Conclusion and Future Work}

In this paper, we have proposed a numerical framework to study self-localized solutions of the Smith's NLSE derived for modelling the nonlinear wave blocking phenomena. The method we have proposed is a spectral renormalization method for the Smith's NLSE and we have showed that self-localized solitons of this model equation can be found for i) constant current velocity gradient, ii) linearly varying current velocity gradient and iii) sinusoidally varying current velocity gradient cases by our approach. By implementing a Fourier spectral scheme with $4^{th}$ order Runge-Kutta time integrator, we have also showed that the self-localized solitons are stable for the cases i and ii, however they are found to be unstable for case iii, for the parameter values considered. Our method can be applied to study soliton blocking by many different analytical and/or experimental current velocity gradients. Additionally, the self-localized solitons of many different extended forms of the Smith's NLSE, such as the ones which include self-steepening terms and/or Raman scattering terms, can be studied within the frame of our approach after some minor modifications. Additionally, the method proposed here can be used to study not only the dynamics of single solitons, but also the dynamics of dual and N-solitons under wave blocking effect.

\end{document}